\pdfoutput=1

\documentclass[11pt,twoside,a4paper]{article}

\usepackage[mono=false]{libertine}
\usepackage[defaultsans]{opensans}
\usepackage{libertinust1math}
\usepackage[T2A]{fontenc}
\usepackage[utf8]{inputenc}
\usepackage[english]{babel}

\usepackage[papersize={200mm,265mm},twoside,
inner=17mm,outer=15mm,top=10mm,bottom=10mm,headsep=4mm,footskip=9mm,
includehead,includefoot,showcrop]{geometry}

\usepackage{amsmath,amsthm,amssymb}

\usepackage{tikz}
\usetikzlibrary{shapes,snakes}
\usepackage{pgfplots}
\usepackage{algpseudocode}
\usepackage{algorithm}
\usepackage{comment}

\usepackage{biblatex}


\theoremstyle{plain}
\newtheorem{theorem}{Theorem}
\newtheorem{lemma}[theorem]{Lemma}

\errorcontextlines\maxdimen

\makeatletter
\newcommand*{\algrule}[1][\algorithmicindent]{\makebox[#1][l]{\hspace*{.5em}\thealgruleextra\vrule height \thealgruleheight depth \thealgruledepth}}%
\newcommand*{\thealgruleextra}{}
\newcommand*{\thealgruleheight}{.75\baselineskip}
\newcommand*{\thealgruledepth}{.25\baselineskip}

\newcount\ALG@printindent@tempcnta
\def\ALG@printindent{%
	\ifnum \theALG@nested>0
	\ifx\ALG@text\ALG@x@notext
	\else
	\unskip
	\addvspace{-1pt}
	\ALG@printindent@tempcnta=1
	\loop
	\algrule[\csname ALG@ind@\the\ALG@printindent@tempcnta\endcsname]%
	\advance \ALG@printindent@tempcnta 1
	\ifnum \ALG@printindent@tempcnta<\numexpr\theALG@nested+1\relax
	\repeat
	\fi
	\fi
}%
\usepackage{etoolbox}
\patchcmd{\ALG@doentity}{\noindent\hskip\ALG@tlm}{\ALG@printindent}{}{\errmessage{failed to patch}}
\makeatother

\newbox\statebox
\newcommand{\myState}[1]{%
	\setbox\statebox=\vbox{#1}%
	\edef\thealgruleheight{\dimexpr \the\ht\statebox+1pt\relax}%
	\edef\thealgruledepth{\dimexpr \the\dp\statebox+1pt\relax}%
	\ifdim\thealgruleheight<.75\baselineskip
	\def\thealgruleheight{\dimexpr .75\baselineskip+1pt\relax}%
	\fi
	\ifdim\thealgruledepth<.25\baselineskip
	\def\thealgruledepth{\dimexpr .25\baselineskip+1pt\relax}%
	\fi
	\State #1%
	\def\thealgruleheight{\dimexpr .75\baselineskip+1pt\relax}%
	\def\thealgruledepth{\dimexpr .25\baselineskip+1pt\relax}%
}

\begin{document}


\title{Backtracking algorithms for constructing the Hamiltonian decomposition of a 4-regular multigraph \thanks{The research is supported by the P.G. Demidov Yaroslavl State University Project VIP-016}}

\author{A. V. Korostil, A. V. Nikolaev \\ 
P.G. Demidov Yaroslavl State University\\
av.korostil@gmail.com, andrei.v.nikolaev@gmail.com
}

\date{}

\maketitle

\begin{abstract}
	We consider a Hamiltonian decomposition problem of partitioning a regular graph into edge-disjoint Hamiltonian cycles. It is known that verifying vertex non-adjacency in the 1-skeleton of the symmetric and asymmetric traveling salesperson polytopes is NP-complete. On the other hand, a sufficient condition for two vertices to be non-adjacent can be formulated as a combinatorial problem of finding a second Hamiltonian decomposition of a 4-regular multigraph. We present two backtracking algorithms for constructing a second Hamiltonian decomposition and verifying vertex non-adjacency: an algorithm based on a simple path extension and an algorithm based on the chain edge fixing procedure. 
	
	Based on the results of computational experiments for undirected multigraphs, both backtracking algorithms lost to the known general variable neighborhood search heuristics. However, for directed multigraphs, the algorithm based on chain fixing of edges showed results comparable to heuristics on instances with an existing solution and better results on infeasible instances where the Hamiltonian decomposition does not exist.
\end{abstract}


\section*{Introduction}

\textit{Hamiltonian decomposition} of a regular multigraph is a partition of its edge set into Hamiltonian cycles.
The problem of finding edge-disjoint Hamiltonian cycles in a given regular graph finds applications in combinatorial optimization \cite{Krar1995}, coding theory \cite{Bae2003, Bail2009}, algorithms for distributed data mining \cite{Clift2002}, analysis of interconnected networks \cite{Hung2011} and other areas.
See also theoretical results on estimating the number of Hamiltonian decompositions of regular graphs \cite{Gleb2017}.
In this paper, the problem of constructing a Hamiltonian decomposition arises from the field of polyhedral combinatorics.

\section{Traveling salesperson polytope}

We consider the classic formulation of the traveling salesperson problem: given a complete weighted graph (or digraph) $K_n = (V,E)$, find a Hamiltonian cycle of minimum weight. Denote by $HC_n$ the set of all Hamiltonian cycles in the graph $K_n$ and assign to each Hamiltonian cycle $x \in HC_n$ the characteristic vector $x^{v} \in \mathbb{R}^E$ according to the following rule:
\[x_e^v = \begin{cases}
	1,& \text{if the cycle $x$ contains an edge $e$},\\
	0,& \text{otherwise}.
\end{cases}\]
Polytope
\[\mathrm{STSP}(n) = \operatorname{conv} \{x^v \ | \ x \in HC_n\}\]
is called \textit{the symmetric traveling salesperson polytope}.

\textit{The asymmetric traveling salesperson polytope} $\mathrm{ATSP}(n)$ is defined similarly as the convex hull of the characteristic vectors of all possible Hamiltonian cycles in the complete digraph $K_n$.

The integer linear programming approach for the traveling salesperson problem was introduced by Dantzig, Fulkerson, and Johnson in their classical work for 49 US cities \cite{Dantz1954}. State-of-the-art exact algorithms for the traveling salesperson problem are based on a partial description of the facets of the traveling salesperson polytope and the branch and cut method for integer linear programming \cite{Appl2006}.

The \textit{1-skeleton} of a polytope is a graph whose vertices are the vertices of the polytope and edges are geometric edges (one-dimensional faces). The study of 1-skeleton is of interest, since, on the one hand, some combinatorial algorithms for such problems as perfect matching, set covering, independent set, object ranking, problems with fuzzy measures, and some others are based on the adjacency relation in a 1-skeleton and the local search technique (when we move from the current solution to the <<best solution>> among adjacent ones) \cite{Aguil2017, Bal1985, Cheg1987, Comb2010, Gab1977}. On the other hand, some characteristics of the 1-skeleton of the problem, such as the diameter and the clique number (the number of vertices in the largest clique), estimate time complexity for various computational models and classes of algorithms \cite{Bond1983,Bond2016,Grot1985}.

Unfortunately, the classical result by Papadimitriou prevents the study of the 1-skeleton of the traveling salesperson polytope.

\begin{theorem}[Papadimitriou \cite{Papa1978}]
	The question of whether two vertices of the polytopes $\mathrm{STSP}(n)$ or $\mathrm{ATSP}(n)$ are non-adjacent is NP-complete.
\end{theorem}

Note that the complementary problem of finding whether two vertices of the 1-skeleton of the traveling salesperson polytope are adjacent is co-NP-complete.

\section{Hamiltonian decomposition and sufficient condition \\ for vertex non-adjacency}

As a result of Papadimitriou's theorem on the NP-completeness of verifying the vertex non-adjacency in the 1-skeleton of the traveling salesperson polytope, sufficient conditions for non-adjacency are of interest.
In particular, polynomial sufficient conditions are known for pyramidal tours \cite{Bond2018}, pyramidal tours with step-backs \cite{Nik2019}, and pedigrees \cite{Arth2006, Arth2013}.
In this paper, we consider the most general of the known -- sufficient condition by Rao \cite{Rao1976}.

Let $x = (V,E_x)$ and $y = (V,E_y)$ be two Hamiltonian cycles on the vertex set $V$. 
We denote by $x \cup y$ a \textit{union multigraph} $(V,E_x \cup E_y)$ that contains a copy of each edge of $x$ and $y$. Note that if two cycles contain the same edge $e$, then both copies of the edge are added to the multigraph $x \cup y$.

\begin{lemma} [Rao \cite{Rao1976}]
	Given two Hamiltonian cycles $x$ and $y$, if the union multigraph $x \cup y$ contains two edge-disjoint Hamiltonian cycles $z$ and $w$ different from $x$ and $y$, then the corresponding vertices $x^v$ and $y^v$ of the traveling salesperson polytope $\mathrm{STSP}(n)$ (or $\mathrm{ATSP}(n)$) are not adjacent.
\end{lemma}

\begin{figure}[h]
	\centering
	\begin{tikzpicture}[scale=1.0]
		\begin{scope}[every node/.style={circle,thick,draw}]
			\node (1) at (0,0) {1};
			\node (2) at (1,1) {2};
			\node (3) at (2.5,1) {3};
			\node (4) at (3.5,0) {4};
			\node (5) at (2.5,-1) {5};
			\node (6) at (1,-1) {6};
		\end{scope}
		\draw [line width=0.3mm] (1) edge (2);
		\draw [line width=0.3mm] (2) edge (3);
		\draw [line width=0.3mm] (3) edge (4);
		\draw [line width=0.3mm] (4) edge (5);
		\draw [line width=0.3mm] (5) edge (6);
		\draw [line width=0.3mm] (6) edge (1);
		\draw (1.75, -1.7) node{\textit{x}};
		
		\begin{scope}[yshift=-4cm]
			\begin{scope}[every node/.style={circle,thick,draw}]
				\node (1) at (0,0) {1};
				\node (2) at (1,1) {2};
				\node (3) at (2.5,1) {3};
				\node (4) at (3.5,0) {4};
				\node (5) at (2.5,-1) {5};
				\node (6) at (1,-1) {6};
			\end{scope}
			\draw [line width=0.3mm] (1) edge (4);
			\draw [line width=0.3mm] (4) edge (6);
			\draw [line width=0.3mm] (6) edge (2);
			\draw [line width=0.3mm] (2) edge (3);
			\draw [line width=0.3mm] (3) edge (5);
			\draw [line width=0.3mm] (5) edge (1);
			\draw (1.75, -1.7) node{\textit{y}};
		\end{scope}
		
		\begin{scope}[xshift=4.5cm,yshift=-2cm]
			\begin{scope}[every node/.style={circle,thick,draw}]
				\node (1) at (0,0) {1};
				\node (2) at (1,1) {2};
				\node (3) at (2.5,1) {3};
				\node (4) at (3.5,0) {4};
				\node (5) at (2.5,-1) {5};
				\node (6) at (1,-1) {6};
			\end{scope}
			\draw [line width=0.3mm] (1) edge (2);
			\draw [line width=0.3mm, bend right=10] (2) edge (3);
			\draw [line width=0.3mm] (3) edge (4);
			\draw [line width=0.3mm] (4) edge (5);
			\draw [line width=0.3mm] (5) edge (6);
			\draw [line width=0.3mm] (6) edge (1);
			\draw [line width=0.3mm] (1) edge (4);
			\draw [line width=0.3mm] (4) edge (6);
			\draw [line width=0.3mm] (6) edge (2);
			\draw [line width=0.3mm, bend left=10] (2) edge (3);
			\draw [line width=0.3mm] (3) edge (5);
			\draw [line width=0.3mm] (5) edge (1);	
			\draw (1.75, -1.7) node{\textit{x $\cup$ y}};
		\end{scope}

		\begin{scope}[xshift=9cm]
			\begin{scope}[every node/.style={circle,thick,draw}]
				\node (1) at (0,0) {1};
				\node (2) at (1,1) {2};
				\node (3) at (2.5,1) {3};
				\node (4) at (3.5,0) {4};
				\node (5) at (2.5,-1) {5};
				\node (6) at (1,-1) {6};
			\end{scope}
			\draw [line width=0.3mm] (1) edge (4);
			\draw [line width=0.3mm] (4) edge (5);
			\draw [line width=0.3mm] (5) edge (3);
			\draw [line width=0.3mm] (3) edge (2);
			\draw [line width=0.3mm] (2) edge (6);
			\draw [line width=0.3mm] (6) edge (1);
			\draw (1.75, -1.7) node{\textit{z}};
			
			\begin{scope}[yshift=-4cm]
				\begin{scope}[every node/.style={circle,thick,draw}]
					\node (1) at (0,0) {1};
					\node (2) at (1,1) {2};
					\node (3) at (2.5,1) {3};
					\node (4) at (3.5,0) {4};
					\node (5) at (2.5,-1) {5};
					\node (6) at (1,-1) {6};
				\end{scope}
				\draw [line width=0.3mm] (1) edge (2);
				\draw [line width=0.3mm] (2) edge (3);
				\draw [line width=0.3mm] (3) edge (4);
				\draw [line width=0.3mm] (4) edge (6);
				\draw [line width=0.3mm] (6) edge (5);
				\draw [line width=0.3mm] (5) edge (1);
				\draw (1.75, -1.7) node{\textit{w}};
			\end{scope}
		\end{scope}
	\end{tikzpicture}
	\caption{An example of a satisfied sufficient condition for non-adjacency}
	\label{image:not_adjacent}
\end{figure}
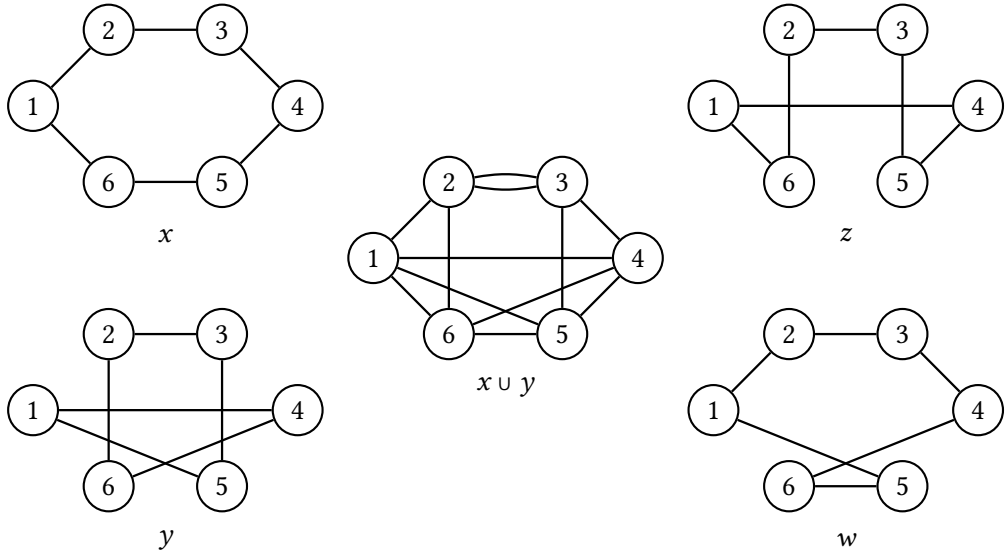

From a geometric point of view, the sufficient condition by Rao means that the segment connecting two vertices $x^v$ and $y^v$ intersects the segment connecting two other vertices $z^v$ and $w^v$ of the traveling salesperson polytope, therefore, they cannot be adjacent.
An example of a satisfied sufficient condition is shown in Fig.~\ref{image:not_adjacent}.

Thus, verifying the vertex non-adjacency in the 1-skeleton of the traveling salesperson polytope can be reduced to finding a Hamiltonian decomposition of the union multigraph $x \cup y$ that is different from the given one.
Let us formulate a sufficient condition for vertex non-adjacency in the form of a combinatorial problem.

\vspace{2mm}

\textbf{Second Hamiltonian decomposition problem.}

\textbf{Instance:} let $x$ and $y$ be two Hamiltonian cycles.

\textbf{Question:} does the union multigraph $x \cup y$ contain a pair of edge-disjoint Hamiltonian cycles $z$ and $w$ different from $x$ and $y$?

\vspace{2mm}

Note that checking whether an arbitrary graph contains a Hamiltonian decomposition is an NP-complete problem already for 4-regular undirected multigraphs and 2-regular directed multigraphs \cite{Per1984}.

Previously, the second Hamiltonian decomposition problem and its application to verifying the vertex non-adjacency in the 1-skeletons of the polytopes $\mathrm{STSP}(n)$ and $\mathrm{ATSP}(n)$ were considered in \cite{Kozl2019, Nik2021}, where several heuristic algorithms were proposed based on constructing vertex-disjoint cycle covers of a graph: simulated annealing and general variable neighborhood search.
Heuristic algorithms have proven to be very efficient on instances that have a solution, especially on undirected graphs. However, in instances that do not have a solution, heuristics face significant difficulties.
In this paper, we consider two exact backtracking algorithms for constructing the second Hamiltonian decomposition of a 4-regular multigraph.

\section{Backtracking algorithm based on simple path extension}

Recall that \textit{backtracking} is one of the general methods for finding a solution to a problem by exhaustive search. The procedure consists of a successive extension of a partial solution. If at the next step such an extension fails, then we backtrack to a shorter partial solution and continue the search further \cite{Skien2008}.

In \cite{Kor2020}, a backtracking algorithm was presented for the problem of finding a second Hamiltonian decomposition of a 4-regular multigraph based on extending a simple path. Below is a modified version of this algorithm.

Let here and below the partial solution consist of two components $z$ and $w$. The idea of the algorithm is to sequentially construct a simple path in the component $z$. Whereas edges not included in $z$ are sent to the component $w$.

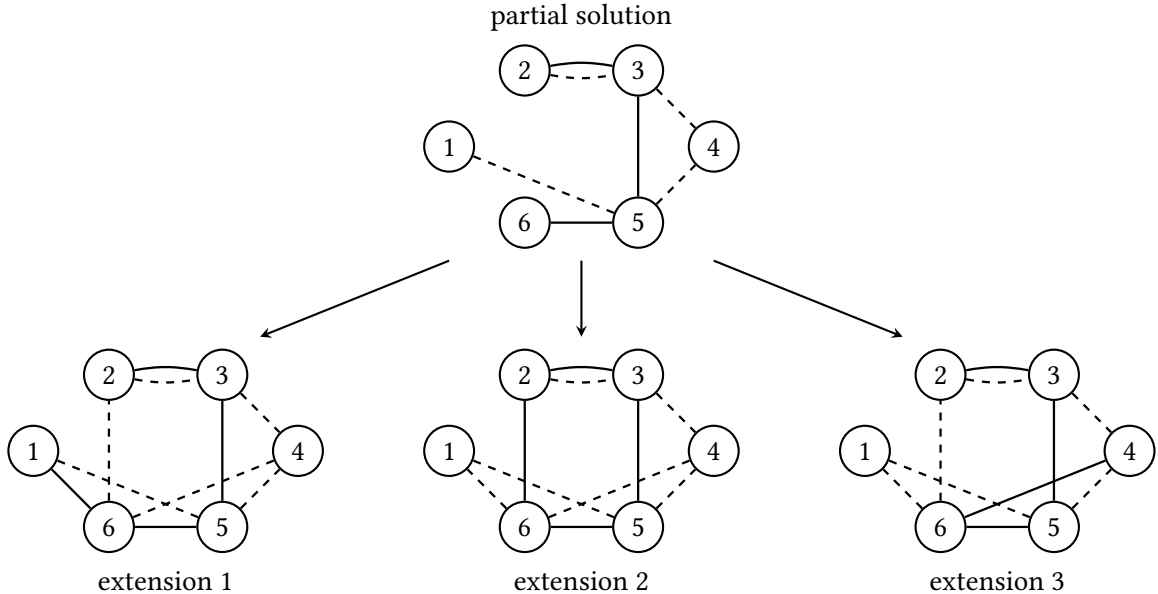
\begin{figure}[h]
	\centering
	\begin{tikzpicture}[scale=1.0]
		\begin{scope}[every node/.style={circle,thick,draw}]
			\node (1) at (0,0) {1};
			\node (2) at (1,1) {2};
			\node (3) at (2.5,1) {3};
			\node (4) at (3.5,0) {4};
			\node (5) at (2.5,-1) {5};
			\node (6) at (1,-1) {6};
		\end{scope}
		\draw [line width=0.3mm, bend left=10] (2) edge (3);
		\draw [line width=0.3mm] (3) edge (5);
		\draw [line width=0.3mm] (5) edge (6);
		\draw [line width=0.3mm, bend right=10, dashed] (2) edge (3);
		\draw [line width=0.3mm, dashed] (3) edge (4);
		\draw [line width=0.3mm, dashed] (5) edge (4);
		\draw [line width=0.3mm, dashed] (5) edge (1);
		\draw (1.75, 1.7) node{partial solution};
		
		\draw [->,>=stealth,thick] (0,-1.5) -- (-2.5,-2.5);
		\draw [->,>=stealth,thick] (1.75,-1.5) -- (1.75,-2.5);
		\draw [->,>=stealth,thick] (3.5,-1.5) -- (6,-2.5);
		
		\begin{scope}[xshift=-5.5cm, yshift=-4cm]
			\begin{scope}[every node/.style={circle,thick,draw}]
				\node (1) at (0,0) {1};
				\node (2) at (1,1) {2};
				\node (3) at (2.5,1) {3};
				\node (4) at (3.5,0) {4};
				\node (5) at (2.5,-1) {5};
				\node (6) at (1,-1) {6};
			\end{scope}
			\draw [line width=0.3mm, bend left=10] (2) edge (3);
			\draw [line width=0.3mm] (3) edge (5);
			\draw [line width=0.3mm] (5) edge (6);
			\draw [line width=0.3mm] (6) edge (1);
			\draw [line width=0.3mm, bend right=10, dashed] (2) edge (3);
			\draw [line width=0.3mm, dashed] (3) edge (4);
			\draw [line width=0.3mm, dashed] (5) edge (4);
			\draw [line width=0.3mm, dashed] (5) edge (1);
			\draw [line width=0.3mm, dashed] (6) edge (2);
			\draw [line width=0.3mm, dashed] (6) edge (4);
			\draw (1.75, -1.7) node{extension 1};
		\end{scope}
		
		\begin{scope}[yshift=-4cm]
			\begin{scope}[every node/.style={circle,thick,draw}]
				\node (1) at (0,0) {1};
				\node (2) at (1,1) {2};
				\node (3) at (2.5,1) {3};
				\node (4) at (3.5,0) {4};
				\node (5) at (2.5,-1) {5};
				\node (6) at (1,-1) {6};
			\end{scope}
			\draw [line width=0.3mm, bend left=10] (2) edge (3);
			\draw [line width=0.3mm] (3) edge (5);
			\draw [line width=0.3mm] (5) edge (6);
			\draw [line width=0.3mm] (6) edge (2);
			\draw [line width=0.3mm, bend right=10, dashed] (2) edge (3);
			\draw [line width=0.3mm, dashed] (3) edge (4);
			\draw [line width=0.3mm, dashed] (5) edge (4);
			\draw [line width=0.3mm, dashed] (5) edge (1);
			\draw [line width=0.3mm, dashed] (6) edge (1);
			\draw [line width=0.3mm, dashed] (6) edge (4);
			\draw (1.75, -1.7) node{extension 2};
		\end{scope}
		
		\begin{scope}[xshift=5.5cm, yshift=-4cm]
			\begin{scope}[every node/.style={circle,thick,draw}]
				\node (1) at (0,0) {1};
				\node (2) at (1,1) {2};
				\node (3) at (2.5,1) {3};
				\node (4) at (3.5,0) {4};
				\node (5) at (2.5,-1) {5};
				\node (6) at (1,-1) {6};
			\end{scope}
			\draw [line width=0.3mm, bend left=10] (2) edge (3);
			\draw [line width=0.3mm] (3) edge (5);
			\draw [line width=0.3mm] (5) edge (6);
			\draw [line width=0.3mm] (6) edge (4);
			\draw [line width=0.3mm, bend right=10, dashed] (2) edge (3);
			\draw [line width=0.3mm, dashed] (3) edge (4);
			\draw [line width=0.3mm, dashed] (5) edge (4);
			\draw [line width=0.3mm, dashed] (5) edge (1);
			\draw [line width=0.3mm, dashed] (6) edge (1);
			\draw [line width=0.3mm, dashed] (6) edge (2);
			\draw (1.75, -1.7) node{extension 3};
		\end{scope}
		
	\end{tikzpicture}
	\caption{Backtracking based on simple path extension}
	\label{image:backtracking_simple_path}
\end{figure}

Consider the example of the $x \cup y$ multigraph shown in Fig.~\ref{image:not_adjacent}. Let us construct a partial solution for it, corresponding to the simple path $2 - 3 - 5 - 6$ (Fig.~\ref{image:backtracking_simple_path}). Here, the edges of the $z$ component are solid, and the edges of the $w$ component are dashed.
Since the degree of each vertex in the multigraph $x \cup y$ is equal to 4, then we can continue a simple path in $z$ from the vertex $6$ in no more than three ways: $(6,1)$, $(6,2)$ and $(6,4)$. Moreover, whichever edge we choose to continue the path, since two edges incident to the vertex $6$ have already been added to $z$, the other two edges can only fall into the component $w$.

We sequentially consider all three options and for each we check the correctness of the partial solution:
\begin{itemize}
	\item extension 1 is invalid because it contains a cycle at vertices $2,3,4,6$ in the component $w$;
	\item extension 2 is invalid because it contains a cycle at vertices $2,3,5,6$ in the component $z$;
	\item extension 3 is correct.
\end{itemize}

In the general case, the correctness conditions for a partial solution have the following form:
\begin{itemize}
	\item $z$ contains a simple path (by construction) or a Hamiltonian cycle;
	\item $w$ contains a forest (a graph without cycles) with vertex degrees at most 2 or a Hamiltonian cycle.
\end{itemize}

If the extension of the partial solution turned out to be incorrect, then we backtrack and consider the next option. If all three extensions are invalid, then this partial solution cannot be extended, and we backtrack to a shorter partial solution.

The general scheme for undirected graphs is presented in the Algorithm~\ref{Alg:backtracking_simple_path}.

\begin{algorithm}[h]
	\caption{Backtracking based on simple path extension}
	\label{Alg:backtracking_simple_path}
	\begin{algorithmic}[1] 
		\Procedure{Backtracking\_Simple\_Path}{$z,w,(i,j),x \cup y$}
		\State Add the edge $(i,j)$ to $z$
		\If{Vertex $i$ in $z$ has 2 incident edges}
		\State Add free edges incident to $i$ to the component $w$
		\EndIf
		\If{partial solution $z,w$ is not correct}
		\State \Return						\Comment{Backtrack to the previous step}
		\EndIf
		\If {$z$ and $w$ are Hamiltonian cycles (different from $x$ and $y$)}
		\State \Return Hamiltonian decomposition of $z$ and $w$ \Comment{Solution found}
		\EndIf
		\State Sort edges $(j,k)$ from $j$ in ascending order of free degrees of vertices $k$
		\For {each free edge $(j,k)$ from the vertex $j$}
		\State $z,w \gets$ \Call{Backtracking\_Simple\_Path}{$z,w,(j,k),x \cup y$}
		\If {Hamiltonian decomposition is found}
		\State \Return Hamiltonian decomposition of $z$ and $w$
		\EndIf
		\EndFor
		\EndProcedure	
		\vspace{2mm}
		\Procedure{Algorithm\_Simple\_Path}{$x \cup y$}
		\State $z,w \gets \emptyset$
		\State Choose an initial edge $(i,j)$ of the multigraph $x \cup y$
		\State $z,w \gets$ \Call{Backtracking\_Simple\_Path}{$z,w,(i,j),x \cup y$}
		\If {$z$ and $w$ are found}
		\State \Return The Hamiltonian decomposition $z$ and $w$ is found (vertices are not adjacent)
		\EndIf
		\State \Return Hamiltonian decomposition not found (vertices are probably adjacent)
		\EndProcedure		
	\end{algorithmic}
\end{algorithm}

The only difference for directed graphs is that the outdegree of each vertex of the multigraph $x \cup y$ is equal to two, hence there are at most two options for extending a simple path in $z$. All other steps are completely similar.

\section{Backtracking algorithm based on chain edge fixing}

The second backtracking algorithm is based on the chain fixation of edges in the components $z$ and $w$.
Algorithms for directed and undirected graphs are slightly different, so we describe them separately.

\subsection{Directed multigraphs}

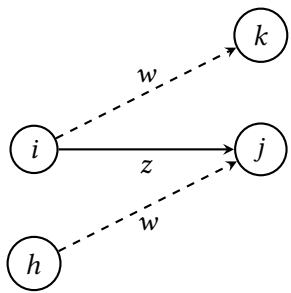
\begin{figure}[h]
	\centering
	\begin{tikzpicture}[scale=1]
		\begin{scope}[every node/.style={circle,thick,draw}]
			\node (i) at (0,0) {$i$};
			\node (j) at (3,0) {$j$};
			\node (k) at (3,1.5) {$k$};
			\node (h) at (0,-1.5) {$h$};
		\end{scope}
		
		\draw [->,>=stealth,thick] (i) edge node[below]{$z$} (j);
		\draw [->,>=stealth,thick,dashed] (i) edge node[above]{$w$} (k);
		\draw [->,>=stealth,thick,dashed] (h) edge node[below]{$w$} (j);
		
	\end{tikzpicture}
	\caption {Fixing the edge $(i,j)$ to $z$}
	\label{Fig_fixed_edges}
\end{figure}

We consider a directed 2-regular multigraph $x \cup y$ for which the indegrees and outdegrees of each vertex are equal to two. Let us choose some edge $(i,j)$ and fix it in the component $z$, then the second edge $(i,k)$ outcoming from $i$ and the second edge $(h,j)$ incoming into $j$ cannot get into $z$. We will fix these edges in $w$ (Fig.~\ref{Fig_fixed_edges}).

The main idea of the algorithm is that the edges $(i,k)$ and $(h,j)$ fixed in $w$, in turn, start recursive chains of fixing edges in $z$, and so on.

\begin{figure}[h]
	\centering
	\begin{tikzpicture}[scale=1.0]
		\begin{scope}[every node/.style={circle,thick,draw}]
			\node (1) at (0,0) {1};
			\node (2) at (1,1) {2};
			\node (3) at (2.5,1) {3};
			\node (4) at (3.5,0) {4};
			\node (5) at (2.5,-1) {5};
			\node (6) at (1,-1) {6};
		\end{scope}
		\draw [->,>=stealth,thick] (1) edge (2);
		\draw [->,>=stealth,thick, bend left=10] (2) edge (3);
		\draw [->,>=stealth,thick] (3) edge (4);
		\draw [->,>=stealth,thick] (4) edge (5);
		\draw [->,>=stealth,thick] (5) edge (6);
		\draw [->,>=stealth,thick] (6) edge (1);
		\draw [->,>=stealth,thick] (1) edge (4);
		\draw [->,>=stealth,thick] (4) edge (6);
		\draw [->,>=stealth,thick] (6) edge (2);
		\draw [->,>=stealth,thick, bend right=10] (2) edge (3);
		\draw [->,>=stealth,thick] (3) edge (5);
		\draw [->,>=stealth,thick] (5) edge (1);	
		\draw (1.75, 1.7) node{$x \cup y$};
		
		\node at (4.75,0) {$\Longrightarrow$};
		
		\begin{scope}[xshift=6cm]
			\begin{scope}[every node/.style={circle,thick,draw}]
				\node (1) at (0,0) {1};
				\node (2) at (1,1) {2};
				\node (3) at (2.5,1) {3};
				\node (4) at (3.5,0) {4};
				\node (5) at (2.5,-1) {5};
				\node (6) at (1,-1) {6};
			\end{scope}
			\draw [->,>=stealth,thick] (1) edge (2);
			\draw [->,>=stealth,thick, bend left=10] (2) edge (3);
			\draw [->,>=stealth,thick] (3) edge (4);
			\draw [->,>=stealth,thick] (4) edge (5);
			\draw [->,>=stealth,thick] (5) edge (6);
			\draw [->,>=stealth,thick] (6) edge (1);
			\draw [->,>=stealth,thick,dashed] (1) edge (4);
			\draw [->,>=stealth,thick,dashed] (4) edge (6);
			\draw [->,>=stealth,thick,dashed] (6) edge (2);
			\draw [->,>=stealth,thick, bend right=10,dashed] (2) edge (3);
			\draw [->,>=stealth,thick,dashed] (3) edge (5);
			\draw [->,>=stealth,thick,dashed] (5) edge (1);	
			\draw (1.75, 1.7) node{$z$ and $w$};
		\end{scope}

	\end{tikzpicture}
	\caption{The result of fixing the edge $(1,2)$ in $z$}
	\label{image:fixing_edges_example}
\end{figure}
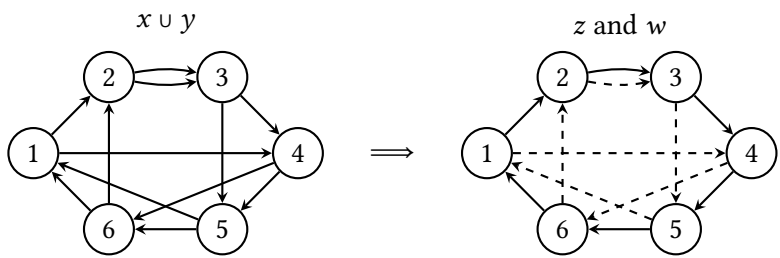

As an example, consider the directed 2-regular multigraph shown in Fig.~\ref{image:fixing_edges_example}.
Note that the multigraph contains two copies of the edge $(2,3)$. Multiple edges cannot get into the same Hamiltonian cycle, so we fix one copy each in $z$ and $w$.
Let's choose some edge, for example $(1,2)$, and fix it in the $z$ component (solid edges), then:
\begin{enumerate}
	\item edge $(1,2)$ is fixed in $z$, hence edges $(1,4)$ and $(6,2)$ go to $w$ (dashed edges);
	\item edge $(1,4)$ is fixed in $w$, hence edge $(3,4)$ goes to $z$, edge $(6,2)$ is fixed in $w$, hence edge $(6,1)$ goes to $z$;
	\item edge $(3,4)$ is fixed in $z$, hence edge $(3,5)$ goes to $w$, edge $(6,1)$ is fixed in $z$, hence edge $(5,1)$ goes to $w$;
	\item edge $(3,5)$ is fixed in $w$, hence edge $(4,5)$ goes to $z$, edge $(5,1)$ is fixed in $w$, hence edge $(5,6)$ goes to $z$.
\end{enumerate}
At the output, given only one edge $(1,2)$ fixed in $z$, we obtain unique Hamiltonian cycles $z$ and $w$ (Fig.~\ref{image:fixing_edges_example}). Considering that the edge $(1,2)$ must belong to at least one Hamiltonian cycle, this 2-regular directed multigraph contains a unique Hamiltonian decomposition, which was found.

The general backtracking scheme for directed multigraphs is given in the pseudocode of the Algorithm~\ref{Alg:backtracking_fixing_directed}.

\begin{algorithm}[p] 
	\caption{Backtracking based on chain edge fixing for directed graphs}
	\label{Alg:backtracking_fixing_directed}
	\begin{algorithmic}[1]
		\Procedure{Chain\_Edge\_Fixing\_Directed}{$(i,j)$ in $z$}		\Comment{Recursive edge fixing}
		\State Fix the edge $(i,j)$ in $z$
		\If{edge $(i,k)$ is not fixed}
		\State \Call{Chain\_Edge\_Fixing\_Directed}{$(i,k)$ in $w$}
		\EndIf
		\If{edge $(h,j)$ is not fixed}
		\State \Call{Chain\_Edge\_Fixing\_Directed}{$(h,j)$ in $w$}
		\EndIf
		\EndProcedure	
		\vspace{2mm}
		\Procedure{Backtracking\_Chain\_Edge\_Fixing\_Directed}{$z,w,(i,j),x \cup y$}
		\State \Call{Chain\_Edge\_Fixing\_Directed}{$(i,j)$ in $z$}
		\If {$z$ and $w$ are Hamiltonian cycles (different from $x$ and $y$)}
		\State \Return $z$ and $w$ 										\Comment{Solution is found}
		\EndIf
		\If {$z$ or $w$ contains a non-Hamiltonian cycle}		\Comment{Partial solution is incorrect}
		\State \Return 			\Comment{Backtrack to the previous step}	
		\EndIf
		\State Select vertex $i'$ with free outcoming edges
		\For {each free edge $(i',j')$ outcoming from the vertex $i'$}
		\State $z,w \gets$ \Call{Backtracking\_Chain\_Edge\_Fixing\_Directed}{$z,w,(i',j'),x \cup y$}
		\If {Hamiltonian decomposition is found}
		\State \Return Hamiltonian decomposition $z$ and $w$
		\EndIf
		\EndFor
		\EndProcedure
		\vspace{2mm}
		\Procedure{Algorithm\_Chain\_Edge\_Fixing\_Directed}{$x \cup y$}
		\State $z,w \gets \emptyset$
		\State Find multiple edges in $x \cup y$ and fix one copy in $z$ and $w$
		\State Choose a free edge $(i,j)$ of the multigraph $x \cup y$
		\State $z,w \gets$ \Call{Backtracking\_Chain\_Edge\_Fixing\_Directed}{$z,w,(i,j),x \cup y$}
		\If {$z$ and $w$ are found}
		\State \Return The Hamiltonian decomposition of $z$ and $w$ is found (vertices are not adjacent)
		\EndIf
		\State \Return Hamiltonian decomposition not found (vertices are probably adjacent)
		\EndProcedure
		\vspace{2mm}		
	\end{algorithmic}
\end{algorithm}

At the data preprocessing step (line 28, Algorithm~\ref{Alg:backtracking_fixing_directed}), we find all multiple edges in the multigraph $x \cup y$ and fix one copy in $z$ and $w$, since these edges cannot get into the same Hamiltonian cycle.

A partial solution is considered correct if the components $z$ and $w$ are directed acyclic graphs with indegrees and outdegrees at most one or directed Hamiltonian cycles. By construction, the indegrees and outdegrees of vertices in $z$ and $w$ cannot be equal to two, so to check the correctness it suffices to verify that there are no non-Hamiltonian cycles.

In contrast to the backtracking algorithm based on the simple path extension (Algorithm~\ref{Alg:backtracking_simple_path}), we choose the vertex $i'$ to extend the partial solution (line 18, Algorithm~\ref{Alg:backtracking_fixing_directed}) in such a way that the chain edge fixing starts from both head and tail of the edge $(i',j')$. The more edges are fixed at one step, the smaller the recursion depth will be.

Note that we do not branch when choosing and fixing the first free edge $(i,j)$ (line 29, Algorithm~\ref{Alg:backtracking_fixing_directed}), since this edge must get into one of the solution components. Let's call the component $z$ the one that contains the edge $(i,j)$.

It should also be noted that although the procedure Chain\_Edge\_Fixing\_Directed (lines 1-9, Algorithm~\ref{Alg:backtracking_fixing_directed}) at each step calls up to two of its subroutines, the total complexity is linear ($O(V)$) since each edge can be fixed at most once, and $|E| = 2|V|$.

\subsection{Undirected multigraphs}

The pseudo-code of the algorithm for undirected multigraphs is presented in Algorithm~\ref{Alg:backtracking_fixing_undirected}.

\begin{algorithm}[h]
	\caption{Backtracking based on chained edge fixing for undirected graphs}
	\label{Alg:backtracking_fixing_undirected}
	\begin{algorithmic}[1]
		\Procedure{Chain\_Edge\_Fixing\_Undirected}{$(i,j)$ in $z$}
		\State Fix the edge $(i,j)$ in $z$
		\If{vertex $i$ in $z$ is incident with 2 fixed edges}
		\State \Call{Chain\_Edge\_Fixing\_Undirected}{$(i,k)$ in $w$}		\Comment{Fix two other edges in $w$}
		\State \Call{Chain\_Edge\_Fixing\_Undirected}{$(i,h)$ in $w$}	
		\EndIf
		\If{vertex $j$ in $z$ is incident with 2 fixed edges}
		\State \Call{Chain\_Edge\_Fixing\_Undirected}{$(j,k)$ in $w$}		\Comment{Fix two other edges in $w$}
		\State \Call{Chain\_Edge\_Fixing\_Undirected}{$(j,h)$ in $w$}	
		\EndIf
		\EndProcedure	
		\vspace{2mm}
		\Procedure{Backtracking\_Chain\_Edge\_Fixing\_Undirected}{$z,w,(i,j),x \cup y$}
		\State \Call{Chain\_Edge\_Fixing\_Undirected}{$(i,j)$ in $z$}
		\If {$z$ and $w$ are Hamiltonian cycles (different from $x$ and $y$)}
		\State \Return $z$ and $w$ 										\Comment{Solution is found}
		\EndIf
		\If {$z$ or $w$ contains a non-Hamiltonian cycle}		\Comment{Partial solution is incorrect}
		\State \Return 			\Comment{Backtrack to the previous step}	
		\EndIf
		\State Choose a vertex $i$ with the minimum degree $d$ of free edges
		\State Sort edges $(i,j_k)$ incident to $i$ in ascending order of free powers $j_k$
		\For {$k \gets 1$ \textbf{to} $d$}
		\State $z,w \gets $\Call{Backtracking\_Chain\_Edge\_Fixing\_Undirected}{$z,w,(i,j_k),x \cup y$}
		\If {Hamiltonian decomposition is found}
		\State \Return Hamiltonian decomposition $z$ and $w$
		\EndIf
		\EndFor
		\EndProcedure		
	\end{algorithmic}
\end{algorithm}

The chain edge fixing procedure is triggered as soon as 2 edges are incident to a vertex in one of the partial solutions, then the other two edges are sent to another partial solution (lines 3-10, Algorithm~\ref{Alg:backtracking_fixing_undirected}).

A partial solution is considered correct if the components $z$ and $w$ are forests with vertex degrees at most two or Hamiltonian cycles. By construction, we add to $z$ and $w$ one edge at a time. Moreover, as soon as the degree of some vertex in one of the components becomes equal to two, then the two remaining edges are sent to another component. Thus, to check the correctness of a partial solution, it suffices to verify that $z$ and $w$ do not contain non-Hamiltonian cycles.

Note that at each step, to expand the partial solution, we choose a vertex $i$ with the minimum degree over free edges in order to reduce the branching factor of the recursion. Edges incident to $i$ are considered in ascending order of degrees of adjacent vertices along free edges (lines 20-27, Algorithm~\ref{Alg:backtracking_fixing_undirected}).

\subsection{Vertex adjacency and data preprocessing}

Let's note that the algorithms considered in this paper were developed for the problem of constructing a Hamiltonian decomposition of a 4-regular multigraph $x \cup y$. The algorithms refer directly to the cycles $x$ and $y$ only when checking that the second decomposition $z$ and $w$ is different from the original one. If we omit this check, then we get algorithms for constructing a Hamiltonian decomposition without any reference to polyhedral combinatorics. Otherwise, if we are interested in exactly the adjacency of vertices in the 1-skeleton of the traveling salesperson polytope, then we can strengthen the considered algorithms by adding a data preprocessing step and checking the known polynomial sufficient conditions for non-adjacency: pyramidal tours \cite{Bond2018}, pyramidal tours with step-backs \cite{Nik2019}, pedigrees \cite{Arth2006, Arth2013}, etc.

\section{Computational experiments}

The algorithms were tested on random directed and undirected 4-regular multigraphs. For each graph size, 100 pairs of random permutations with a uniform probability distribution were generated using the Fisher-Yates shuffle algorithm \cite{Knuth1997}.

For comparison, we chose 3 algorithms:
\begin{itemize}
	\item backtracking based on simple path extension (BSP);
	\item backtracking based on chain edge fixing (BCEF);
	\item general variable neighborhood search (GVNS) heuristic algorithm from \cite{Nik2021}, which is a modification of the simulated annealing algorithm \cite{Kozl2019} and is based on constructing a vertex-disjoint cycle covers by reduction to perfect matching and several cycle merging operations.
\end{itemize}

Backtracking algorithms are implemented in Python, for the general variable neighborhood search heuristics, a ready-made implementation in Node.js \cite{Nik2021} is taken. Computational experiments were carried out on an Intel(R) Core(TM) i5-4460 machine with a 3.20GHz CPU and 16GB RAM.

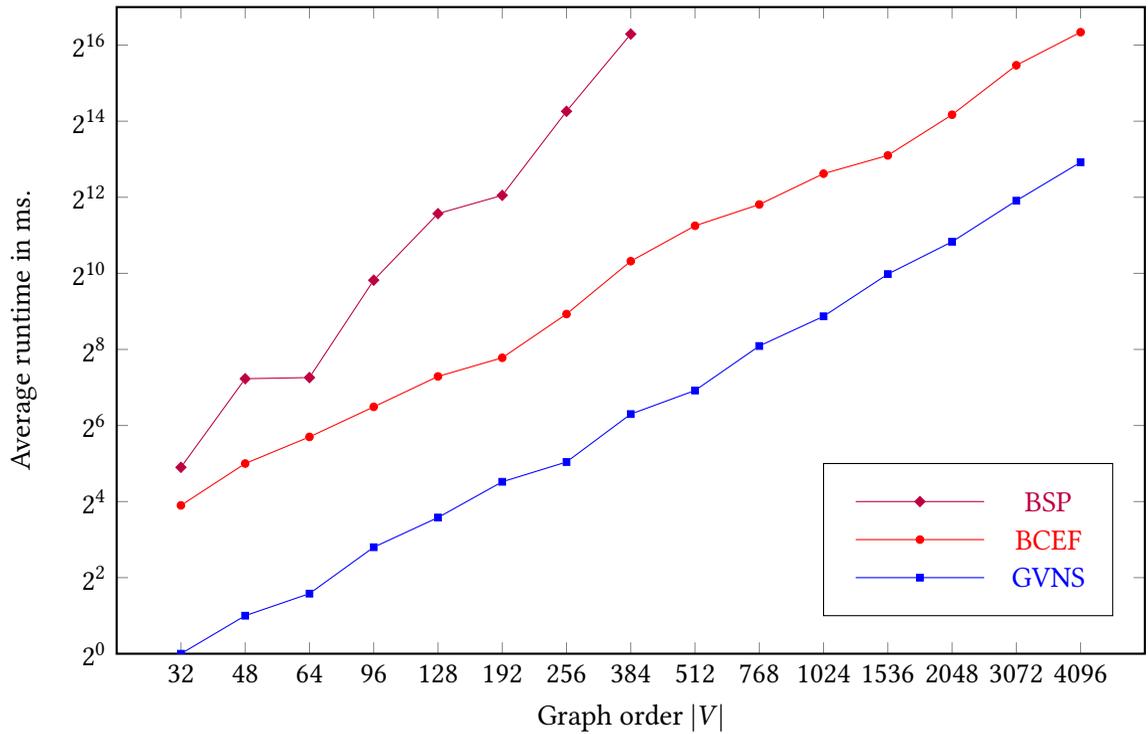
\begin{figure}[p]
	\centering
	\begin{tikzpicture}[scale=1]
		\begin{axis}[
			y=0.5cm,
			x=0.85cm,
			axis line style = thick,
			xlabel={Graph order $|V|$},
			ylabel={Average runtime in ms.},
			xtick       = {1,2,3,4,5,6,7,8,9,10,11,12,13,14,15},
			xticklabels = {32,48,64,96,128,192,256,384,512,768,1024,1536,2048,3072,4096},
			ytick       = {0,2,4,6,8,10,12,14,16},
			yticklabels = {$2^{0}$,$2^{2}$,$2^{4}$,$2^{6}$,$2^{8}$,$2^{10}$,$2^{12}$,$2^{14}$,$2^{16}$},
			xmin=0,
			xmax=16,
			ymin=0,
			ymax=17.]
			
			\node [diamond,draw,fill,inner sep=1pt,purple] (BSP1) at (axis cs: 1,4.9) {};
			\node [diamond,draw,fill,inner sep=1pt,purple] (BSP2) at (axis cs: 2,7.23) {};
			\node [diamond,draw,fill,inner sep=1pt,purple] (BSP3) at (axis cs: 3,7.26) {};
			\node [diamond,draw,fill,inner sep=1pt,purple] (BSP4) at (axis cs: 4,9.82) {};
			\node [diamond,draw,fill,inner sep=1pt,purple] (BSP5) at (axis cs: 5,11.57) {};
			\node [diamond,draw,fill,inner sep=1pt,purple] (BSP6) at (axis cs: 6,12.05) {};
			\node [diamond,draw,fill,inner sep=1pt,purple] (BSP7) at (axis cs: 7,14.26) {};
			\node [diamond,draw,fill,inner sep=1pt,purple] (BSP8) at (axis cs: 8,16.29) {};
			
			\draw [purple] (BSP1) -- (BSP2) -- (BSP3) -- (BSP4) -- (BSP5) -- (BSP6) -- (BSP7) -- (BSP8);
			
			\node [circle,draw,fill,inner sep=1pt,red] (BCEF1) at (axis cs: 1,3.9) {};
			\node [circle,draw,fill,inner sep=1pt,red] (BCEF2) at (axis cs: 2,5) {};
			\node [circle,draw,fill,inner sep=1pt,red] (BCEF3) at (axis cs: 3,5.7) {};
			\node [circle,draw,fill,inner sep=1pt,red] (BCEF4) at (axis cs: 4,6.49) {};
			\node [circle,draw,fill,inner sep=1pt,red] (BCEF5) at (axis cs: 5,7.29) {};
			\node [circle,draw,fill,inner sep=1pt,red] (BCEF6) at (axis cs: 6,7.78) {};
			\node [circle,draw,fill,inner sep=1pt,red] (BCEF7) at (axis cs: 7,8.93) {};
			\node [circle,draw,fill,inner sep=1pt,red] (BCEF8) at (axis cs: 8,10.32) {};
			\node [circle,draw,fill,inner sep=1pt,red] (BCEF9) at (axis cs: 9,11.25) {};
			\node [circle,draw,fill,inner sep=1pt,red] (BCEF10) at (axis cs: 10,11.81) {};
			\node [circle,draw,fill,inner sep=1pt,red] (BCEF11) at (axis cs: 11,12.62) {};
			\node [circle,draw,fill,inner sep=1pt,red] (BCEF12) at (axis cs: 12,13.1) {};
			\node [circle,draw,fill,inner sep=1pt,red] (BCEF13) at (axis cs: 13,14.17) {};
			\node [circle,draw,fill,inner sep=1pt,red] (BCEF14) at (axis cs: 14,15.47) {};
			\node [circle,draw,fill,inner sep=1pt,red] (BCEF15) at (axis cs: 15,16.34) {};
			
			\draw [red] (BCEF1) -- (BCEF2) -- (BCEF3) -- (BCEF4) -- (BCEF5) -- (BCEF6) -- (BCEF7) -- (BCEF8) -- (BCEF9) -- (BCEF10) -- (BCEF11) -- (BCEF12) -- (BCEF13) -- (BCEF14) -- (BCEF15);

			\node [draw,fill,inner sep=1.25pt,blue] (GVNS1) at (axis cs: 1,0) {};
			\node [draw,fill,inner sep=1.25pt,blue] (GVNS2) at (axis cs: 2,1) {};
			\node [draw,fill,inner sep=1.25pt,blue] (GVNS3) at (axis cs: 3,1.58) {};
			\node [draw,fill,inner sep=1.25pt,blue] (GVNS4) at (axis cs: 4,2.8) {};
			\node [draw,fill,inner sep=1.25pt,blue] (GVNS5) at (axis cs: 5,3.58) {};
			\node [draw,fill,inner sep=1.25pt,blue] (GVNS6) at (axis cs: 6,4.52) {};
			\node [draw,fill,inner sep=1.25pt,blue] (GVNS7) at (axis cs: 7,5.04) {};
			\node [draw,fill,inner sep=1.25pt,blue] (GVNS8) at (axis cs: 8,6.3) {};
			\node [draw,fill,inner sep=1.25pt,blue] (GVNS9) at (axis cs: 9,6.92) {};
			\node [draw,fill,inner sep=1.25pt,blue] (GVNS10) at (axis cs: 10,8.09) {};
			\node [draw,fill,inner sep=1.25pt,blue] (GVNS11) at (axis cs: 11,8.87) {};
			\node [draw,fill,inner sep=1.25pt,blue] (GVNS12) at (axis cs: 12,9.98) {};
			\node [draw,fill,inner sep=1.25pt,blue] (GVNS13) at (axis cs: 13,10.83) {};
			\node [draw,fill,inner sep=1.25pt,blue] (GVNS14) at (axis cs: 14,11.91) {};
			\node [draw,fill,inner sep=1.25pt,blue] (GVNS15) at (axis cs: 15,12.92) {};
			
			\draw [blue] (GVNS1) -- (GVNS2) -- (GVNS3) -- (GVNS4) -- (GVNS5) -- (GVNS6) -- (GVNS7) -- (GVNS8) -- (GVNS9) -- (GVNS10) -- (GVNS11) -- (GVNS12) -- (GVNS13) -- (GVNS14) -- (GVNS15);

			\draw [purple] (axis cs: 11.5,4) -- (axis cs: 13.5,4);
			\node [diamond,draw,fill,inner sep=1pt,purple] at (axis cs: 12.5,4) {};
			\node [purple] at (axis cs: 14.5,4) {BSP};
			
			\draw [red] (axis cs: 11.5,3) -- (axis cs: 13.5,3);
			\node [circle,draw,fill,inner sep=1pt,red] at (axis cs: 12.5,3) {};
			\node [red] at (axis cs: 14.5,3) {BCEF};
			
			\draw [blue] (axis cs: 11.5,2) -- (axis cs: 13.5,2);
			\node [draw,fill,inner sep=1.25pt,blue] at (axis cs: 12.5,2) {};
			\node [blue] at (axis cs: 14.5,2) {GVNS};
			
			\draw (axis cs: 11,5) -- (axis cs: 15.5,5) -- (axis cs: 15.5,1) -- (axis cs: 11,1) -- cycle;
			
		\end{axis}
	\end{tikzpicture}
	\caption{Computational results for undirected graphs}
	\label{image:computational_results_graph_undirected}
\end{figure}

\begin{figure}[p]
	\centering
	\begin{tikzpicture}[scale=1]
		\begin{axis}[
			y=0.5cm,
			x=0.85cm,
			axis line style = thick,
			xlabel={Graph order $|V|$},
			ylabel={Average runtime in ms.},
			xtick       = {1,2,3,4,5,6,7,8,9,10,11,12,13,14,15},
			xticklabels = {32,48,64,96,128,192,256,384,512,768,1024,1536,2048,3072,4096},
			ytick       = {0,2,4,6,8,10,12,14},
			yticklabels = {$2^{0}$,$2^{2}$,$2^{4}$,$2^{6}$,$2^{8}$,$2^{10}$,$2^{12}$,$2^{14}$},
			xmin=0,
			xmax=16,
			ymin=0,
			ymax=16.]
			
			\node [diamond,draw,fill,inner sep=1pt,purple] (BSP1) at (axis cs: 1,6.1) {};
			\node [diamond,draw,fill,inner sep=1pt,purple] (BSP2) at (axis cs: 2,10.84) {};
			\node [diamond,draw,fill,inner sep=1pt,purple] (BSP3) at (axis cs: 3,15.17) {};
			
			\draw [purple] (BSP1) -- (BSP2) -- (BSP3);
			
			\node [circle,draw,fill,inner sep=1pt,red] (BCEF1) at (axis cs: 1,2.58) {};
			\node [circle,draw,fill,inner sep=1pt,red] (BCEF2) at (axis cs: 2,3.46) {};
			\node [circle,draw,fill,inner sep=1pt,red] (BCEF3) at (axis cs: 3,4.7) {};
			\node [circle,draw,fill,inner sep=1pt,red] (BCEF4) at (axis cs: 4,5) {};
			\node [circle,draw,fill,inner sep=1pt,red] (BCEF5) at (axis cs: 5,5.78) {};
			\node [circle,draw,fill,inner sep=1pt,red] (BCEF6) at (axis cs: 6,5.88) {};
			\node [circle,draw,fill,inner sep=1pt,red] (BCEF7) at (axis cs: 7,7.06) {};
			\node [circle,draw,fill,inner sep=1pt,red] (BCEF8) at (axis cs: 8,6.37) {};
			\node [circle,draw,fill,inner sep=1pt,red] (BCEF9) at (axis cs: 9,7.16) {};
			\node [circle,draw,fill,inner sep=1pt,red] (BCEF10) at (axis cs: 10,8.73) {};
			\node [circle,draw,fill,inner sep=1pt,red] (BCEF11) at (axis cs: 11,9.72) {};
			\node [circle,draw,fill,inner sep=1pt,red] (BCEF12) at (axis cs: 12,8.52) {};
			\node [circle,draw,fill,inner sep=1pt,red] (BCEF13) at (axis cs: 13,10.69) {};
			\node [circle,draw,fill,inner sep=1pt,red] (BCEF14) at (axis cs: 14,10.04) {};
			\node [circle,draw,fill,inner sep=1pt,red] (BCEF15) at (axis cs: 15,10.59) {};
			
			\draw [red] (BCEF1) -- (BCEF2) -- (BCEF3) -- (BCEF4) -- (BCEF5) -- (BCEF6) -- (BCEF7) -- (BCEF8) -- (BCEF9) -- (BCEF10) -- (BCEF11) -- (BCEF12) -- (BCEF13) -- (BCEF14) -- (BCEF15);

			\node [draw,fill,inner sep=1.25pt,blue] (GVNS1) at (axis cs: 1,0) {};
			\node [draw,fill,inner sep=1.25pt,blue] (GVNS2) at (axis cs: 2,1) {};
			\node [draw,fill,inner sep=1.25pt,blue] (GVNS3) at (axis cs: 3,1) {};
			\node [draw,fill,inner sep=1.25pt,blue] (GVNS4) at (axis cs: 4,2.32) {};
			\node [draw,fill,inner sep=1.25pt,blue] (GVNS5) at (axis cs: 5,3.17) {};
			\node [draw,fill,inner sep=1.25pt,blue] (GVNS6) at (axis cs: 6,3.8) {};
			\node [draw,fill,inner sep=1.25pt,blue] (GVNS7) at (axis cs: 7,6.53) {};
			\node [draw,fill,inner sep=1.25pt,blue] (GVNS8) at (axis cs: 8,6.7) {};
			\node [draw,fill,inner sep=1.25pt,blue] (GVNS9) at (axis cs: 9,6.8) {};
			\node [draw,fill,inner sep=1.25pt,blue] (GVNS10) at (axis cs: 10,8.65) {};
			\node [draw,fill,inner sep=1.25pt,blue] (GVNS11) at (axis cs: 11,10.71) {};
			\node [draw,fill,inner sep=1.25pt,blue] (GVNS12) at (axis cs: 12,9.88) {};
			\node [draw,fill,inner sep=1.25pt,blue] (GVNS13) at (axis cs: 13,10.95) {};
			\node [draw,fill,inner sep=1.25pt,blue] (GVNS14) at (axis cs: 14,11.55) {};
			\node [draw,fill,inner sep=1.25pt,blue] (GVNS15) at (axis cs: 15,12.24) {};
			
			\draw [blue] (GVNS1) -- (GVNS2) -- (GVNS3) -- (GVNS4) -- (GVNS5) -- (GVNS6) -- (GVNS7) -- (GVNS8) -- (GVNS9) -- (GVNS10) -- (GVNS11) -- (GVNS12) -- (GVNS13) -- (GVNS14) -- (GVNS15);

			\draw [purple] (axis cs: 11.5,4) -- (axis cs: 13.5,4);
			\node [diamond,draw,fill,inner sep=1pt,purple] at (axis cs: 12.5,4) {};
			\node [purple] at (axis cs: 14.5,4) {BSP};
			
			\draw [red] (axis cs: 11.5,3) -- (axis cs: 13.5,3);
			\node [circle,draw,fill,inner sep=1pt,red] at (axis cs: 12.5,3) {};
			\node [red] at (axis cs: 14.5,3) {BCEF};
			
			\draw [blue] (axis cs: 11.5,2) -- (axis cs: 13.5,2);
			\node [draw,fill,inner sep=1.25pt,blue] at (axis cs: 12.5,2) {};
			\node [blue] at (axis cs: 14.5,2) {GVNS};
			
			\draw (axis cs: 11,5) -- (axis cs: 15.5,5) -- (axis cs: 15.5,1) -- (axis cs: 11,1) -- cycle;
			
		\end{axis}
	\end{tikzpicture}
	\caption{Computational results for feasible directed graphs}
	\label{image:computational_results_graph_directed_solution}
\end{figure}
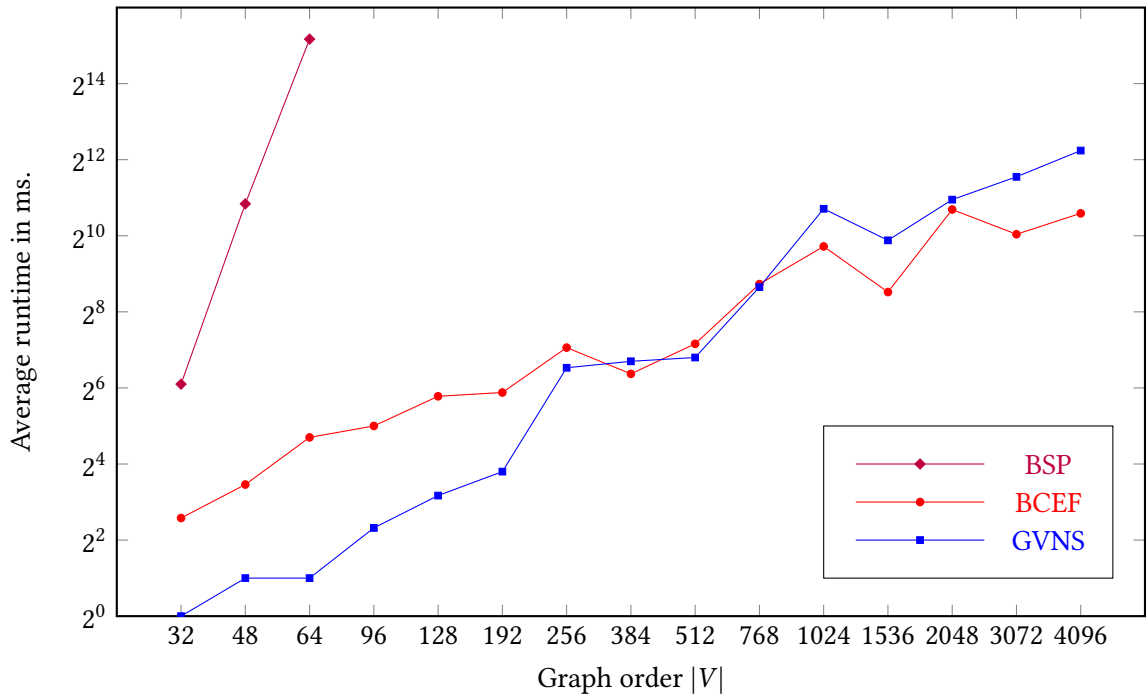

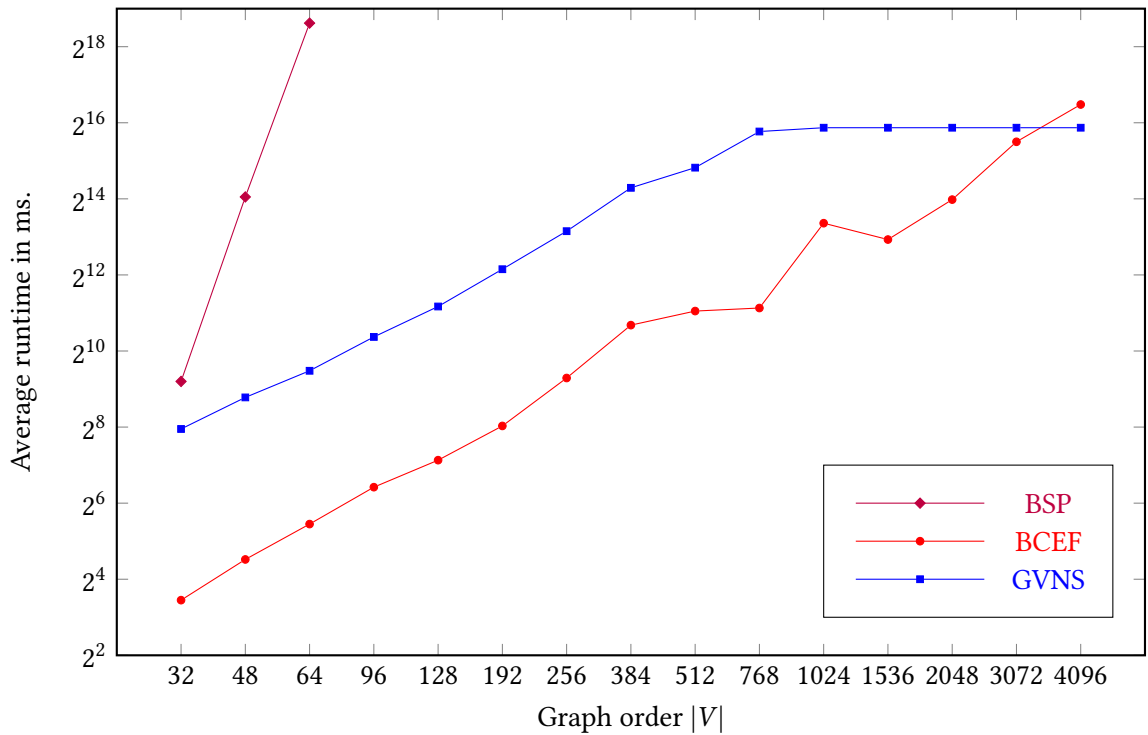
\begin{figure}[p]
	\centering
	\begin{tikzpicture}[scale=1]
		\begin{axis}[
			y=0.5cm,
			x=0.85cm,
			axis line style = thick,
			xlabel={Graph order $|V|$},
			ylabel={Average runtime in ms.},
			xtick       = {1,2,3,4,5,6,7,8,9,10,11,12,13,14,15},
			xticklabels = {32,48,64,96,128,192,256,384,512,768,1024,1536,2048,3072,4096},
			ytick       = {2,4,6,8,10,12,14,16,18},
			yticklabels = {$2^{2}$,$2^{4}$,$2^{6}$,$2^{8}$,$2^{10}$,$2^{12}$,$2^{14}$,$2^{16}$,$2^{18}$},
			xmin=0,
			xmax=16,
			ymin=2,
			ymax=19.]
			
			\node [diamond,draw,fill,inner sep=1pt,purple] (BSP1) at (axis cs: 1,9.2) {};
			\node [diamond,draw,fill,inner sep=1pt,purple] (BSP2) at (axis cs: 2,14.05) {};
			\node [diamond,draw,fill,inner sep=1pt,purple] (BSP3) at (axis cs: 3,18.62) {};
			
			\draw [purple] (BSP1) -- (BSP2) -- (BSP3);
			
			\node [circle,draw,fill,inner sep=1pt,red] (BCEF1) at (axis cs: 1,3.45) {};
			\node [circle,draw,fill,inner sep=1pt,red] (BCEF2) at (axis cs: 2,4.52) {};
			\node [circle,draw,fill,inner sep=1pt,red] (BCEF3) at (axis cs: 3,5.45) {};
			\node [circle,draw,fill,inner sep=1pt,red] (BCEF4) at (axis cs: 4,6.42) {};
			\node [circle,draw,fill,inner sep=1pt,red] (BCEF5) at (axis cs: 5,7.13) {};
			\node [circle,draw,fill,inner sep=1pt,red] (BCEF6) at (axis cs: 6,8.03) {};
			\node [circle,draw,fill,inner sep=1pt,red] (BCEF7) at (axis cs: 7,9.29) {};
			\node [circle,draw,fill,inner sep=1pt,red] (BCEF8) at (axis cs: 8,10.68) {};
			\node [circle,draw,fill,inner sep=1pt,red] (BCEF9) at (axis cs: 9,11.05) {};
			\node [circle,draw,fill,inner sep=1pt,red] (BCEF10) at (axis cs: 10,11.13) {};
			\node [circle,draw,fill,inner sep=1pt,red] (BCEF11) at (axis cs: 11,13.36) {};
			\node [circle,draw,fill,inner sep=1pt,red] (BCEF12) at (axis cs: 12,12.93) {};
			\node [circle,draw,fill,inner sep=1pt,red] (BCEF13) at (axis cs: 13,13.98) {};
			\node [circle,draw,fill,inner sep=1pt,red] (BCEF14) at (axis cs: 14,15.5) {};
			\node [circle,draw,fill,inner sep=1pt,red] (BCEF15) at (axis cs: 15,16.48) {};
			
			\draw [red] (BCEF1) -- (BCEF2) -- (BCEF3) -- (BCEF4) -- (BCEF5) -- (BCEF6) -- (BCEF7) -- (BCEF8) -- (BCEF9) -- (BCEF10) -- (BCEF11) -- (BCEF12) -- (BCEF13) -- (BCEF14) -- (BCEF15);

			\node [draw,fill,inner sep=1.25pt,blue] (GVNS1) at (axis cs: 1,7.95) {};
			\node [draw,fill,inner sep=1.25pt,blue] (GVNS2) at (axis cs: 2,8.78) {};
			\node [draw,fill,inner sep=1.25pt,blue] (GVNS3) at (axis cs: 3,9.48) {};
			\node [draw,fill,inner sep=1.25pt,blue] (GVNS4) at (axis cs: 4,10.37) {};
			\node [draw,fill,inner sep=1.25pt,blue] (GVNS5) at (axis cs: 5,11.17) {};
			\node [draw,fill,inner sep=1.25pt,blue] (GVNS6) at (axis cs: 6,12.15) {};
			\node [draw,fill,inner sep=1.25pt,blue] (GVNS7) at (axis cs: 7,13.15) {};
			\node [draw,fill,inner sep=1.25pt,blue] (GVNS8) at (axis cs: 8,14.29) {};
			\node [draw,fill,inner sep=1.25pt,blue] (GVNS9) at (axis cs: 9,14.82) {};
			\node [draw,fill,inner sep=1.25pt,blue] (GVNS10) at (axis cs: 10,15.77) {};
			\node [draw,fill,inner sep=1.25pt,blue] (GVNS11) at (axis cs: 11,15.87) {};
			\node [draw,fill,inner sep=1.25pt,blue] (GVNS12) at (axis cs: 12,15.87) {};
			\node [draw,fill,inner sep=1.25pt,blue] (GVNS13) at (axis cs: 13,15.87) {};
			\node [draw,fill,inner sep=1.25pt,blue] (GVNS14) at (axis cs: 14,15.87) {};
			\node [draw,fill,inner sep=1.25pt,blue] (GVNS15) at (axis cs: 15,15.87) {};
			
			\draw [blue] (GVNS1) -- (GVNS2) -- (GVNS3) -- (GVNS4) -- (GVNS5) -- (GVNS6) -- (GVNS7) -- (GVNS8) -- (GVNS9) -- (GVNS10) -- (GVNS11) -- (GVNS12) -- (GVNS13) -- (GVNS14) -- (GVNS15);

			\draw [purple] (axis cs: 11.5,6) -- (axis cs: 13.5,6);
			\node [diamond,draw,fill,inner sep=1pt,purple] at (axis cs: 12.5,6) {};
			\node [purple] at (axis cs: 14.5,6) {BSP};
			
			\draw [red] (axis cs: 11.5,5) -- (axis cs: 13.5,5);
			\node [circle,draw,fill,inner sep=1pt,red] at (axis cs: 12.5,5) {};
			\node [red] at (axis cs: 14.5,5) {BCEF};
			
			\draw [blue] (axis cs: 11.5,4) -- (axis cs: 13.5,4);
			\node [draw,fill,inner sep=1.25pt,blue] at (axis cs: 12.5,4) {};
			\node [blue] at (axis cs: 14.5,4) {GVNS};
			
			\draw (axis cs: 11,7) -- (axis cs: 15.5,7) -- (axis cs: 15.5,3) -- (axis cs: 11,3) -- cycle;
			
		\end{axis}
	\end{tikzpicture}
	\caption{Computational results for infeasible directed graphs}
	\label{image:computational_results_graph_directed_no_solution}
\end{figure}

The results of computational experiments for undirected graphs are presented in the Table~\ref{table:random_undirected} and in Fig.~\ref{image:computational_results_graph_undirected}. Results for directed graphs are presented in Table~\ref{table:random_directed} and in Fig.~\ref{image:computational_results_graph_directed_solution} and Fig.~\ref{image:computational_results_graph_directed_no_solution}.

\begin{table}[p]
	\centering
	\caption{Computational results for 100 random undirected Hamiltonian cycles}
	\label{table:random_undirected}
	\resizebox{\textwidth}{!}{%
		\begin{tabular}{|*{13}{r|}}
			\hline
			& \multicolumn{4}{c|}{BSP} &  \multicolumn{4}{c|}{BCEF} & \multicolumn{4}{c|}{GVNS} \\ 
			\hline
			& \multicolumn{2}{c|}{Feasible} & \multicolumn{2}{c|}{Infeasible} & \multicolumn{2}{c|}{Feasible} & \multicolumn{2}{c|}{Infeasible} & \multicolumn{2}{c|}{Feasible} & \multicolumn{2}{c|}{<<Infeasible>>} \\ \hline
			|V| & N & time (s) & N & times (s) & N & time (s) & N & time (s) & N & time (s) & N & time (s)  \\
			\hline
			32 & $100$ & $0.030$ & $-$ & $-$ & $100$ & $0.015$ & $-$ & $-$ & $100$ & $0.001$ & $-$ & $-$  \\ 
			\hline
			48 & $100$ & $0.151$ & $-$ & $-$ & $100$ & $0.032$ & $-$ & $-$ & $100$ & $0.002$ & $-$ & $-$ \\ 
			\hline
			64 & $100$ & $0.154$ & $-$ & $-$ & $100$ & $0.052$ & $-$ & $-$ & $100$ & $0.003$ & $-$ & $-$ \\
			\hline
			96 & $100$ & $0.908$ & $-$ & $-$ & $100$ & $0.09$ & $-$ & $-$ & $100$ & $0.007$ & $-$ & $-$ \\
			\hline
			128 & $100$ & $3.045$ & $-$ & $-$ & $100$ & $0.157$ & $-$ & $-$ & $100$ & $0.012$ & $-$ & $-$\\
			\hline
			192 & $100$ & $4.257$ & $-$ & $-$ & $100$ & $0.22$ & $-$ & $-$ & $100$ & $0.023$ & $-$ & $-$\\
			\hline
			256 & $100$ & $19.625$ & $-$ & $-$ & $100$ & $0.487$ & $-$ & $-$ & $100$ & $0.033$ & $-$ & $-$\\ 
			\hline
			384 & $9$ & $80.037$ & $-$ & $-$ & $100$ & $1.278$ & $-$ & $-$ & $100$ & $0.079$ & $-$ & $-$\\
			\hline
			512 & $0$ & $-$ & $-$ & $-$ & $100$ & $2.447$ & $-$ & $-$ & $100$ & $0.121$ & $-$ & $-$\\
			\hline
			768 & $0$ & $-$ & $-$ & $-$ & $100$ & $3.591$ & $-$ & $-$ & $100$ & $0.272$ & $-$ & $-$\\
			\hline
			1024 & $0$ & $-$ & $-$ & $-$ & $100$ & $6.305$ & $-$ & $-$ & $100$ & $0.468$ & $-$ & $-$\\
			\hline
			1536 & $0$ & $-$ & $-$ & $-$ & $100$ & $8.799$ & $-$ & $-$ & $100$ & $1.011$ & $-$ & $-$\\
			\hline
			2048 & $0$ & $-$ & $-$ & $-$ & $100$ & $18.499$ & $-$ & $-$ & $100$ & $1.821$ & $-$ & $-$\\
			\hline
			3072 & $0$ & $-$ & $-$ & $-$ & $100$ & $45.542$ & $-$ & $-$ & $100$ & $3.852$ & $-$ & $-$\\
			\hline
			4096 & $0$ & $-$ & $-$ & $-$ & $86$ & $82.973$ & $-$ & $-$ & $100$ & $7.768$ & $-$ & $-$\\
			\hline
		\end{tabular}
	}
\end{table}

For each set of 100 tests, the average running time of the algorithms in seconds is given separately for feasible and infeasible instances. 
For two backtracking algorithms (BSP and BCEF), a time limit of 2 hours was set for each set of 100 test problems. Accordingly, the tables indicate how many instances out of 100 the algorithm managed to solve in 2 hours. The general variable neighborhood search heuristics (GVNS) was set to a limit of 1 minute per instance. This is because the heuristic algorithm has a one-sided error. If the algorithm finds a solution, then the instance is feasible. However, the heuristic algorithm cannot guarantee that the problem is infeasible, only that the solution is not found in a given time or number of iterations.

It is known that random undirected regular graphs have a Hamiltonian decomposition with a very high probability \cite{Kim2001}. Indeed, for all test problems on undirected multigraphs (Table~\ref{table:random_undirected}), there was a second Hamiltonian decomposition into cycles different from the original ones, and the vertices of the traveling salesperson polytope were not adjacent. From a geometric point of view, this means that the degrees of vertices in a 1-skeleton are much less than the total number of vertices, so two random vertices are not adjacent with a high probability.

According to the computational results for undirected Hamiltonian cycles, both backtracking algorithms lost out to the general variable neighborhood search heuristics (GVNS). In the allotted time, the algorithm based on simple path extension (BSP) solved 709 instances, the algorithm based on chain edge fixing (BCEF) solved 1486 instances, and the general variable neighborhood search (GVNS) solved all 1500 out of 1500 instances. GVNS was on average 13 times faster than BCEF and 640 times faster than BSP. Among the two backtracking algorithms, the chain edge fixing (BCEF) algorithm proved to be significantly more efficient, solving all test instances up to 3072 vertices and showing the running time on the solved problems on average 48 times faster than the simple path construction (BSP).

\begin{table}[t]
	\centering
	\caption{Computational results for 100 random directed Hamiltonian cycles}
	\label{table:random_directed}
	\resizebox{\textwidth}{!}{%
		\begin{tabular}{|*{13}{r|}}
			\hline
			& \multicolumn{4}{c|}{BSP} &  \multicolumn{4}{c|}{BCEF} & \multicolumn{4}{c|}{GVNS} \\ 
			\hline
			& \multicolumn{2}{c|}{Feasible} & \multicolumn{2}{c|}{Infeasible} & \multicolumn{2}{c|}{Feasible} & \multicolumn{2}{c|}{Infeasible} & \multicolumn{2}{c|}{Feasible} & \multicolumn{2}{c|}{<<Infeasible>>} \\ \hline
			|V| & N & times (s) & N & time (s) & N & time (s) & N & time (s) & N & time (s) & N & time (s)  \\
			\hline
			32 & $30$ & $0.069$ & $70$ & $0.592$ & $30$ & $0.006$ & $70$ & $0.011$ & $30$ & $0.001$ & $70$ & $0.248$ \\ 
			\hline
			48 & $20$ & $1.838$ & $80$ & $17.005$ & $20$ & $0.011$ & $80$ & $0.023$ & $20$ & $0.002$ & $80$ & $0.440$ \\ 
			\hline
			64 & $8$ & $36.850$ & $17$ & $403.016$ & $20$ & $0.026$ & $80$ & $0.044$ & $20$ & $0.002$ & $80$ & $0.714$\\
			\hline
			96 & $0$ & $-$ & $0$ & $-$ & $19$ & $0.032$ & $81$ & $0.086$ & $19$ & $0.005$ & $81$ & $1.330$ \\
			\hline
			128 & $-$ & $-$ & $-$ & $-$ & $18$ & $0.055$ & $82$ & $0.141$ & $18$ & $0.009$ & $82$ & $2.312$ \\
			\hline
			192 & $-$ & $-$ & $-$ & $-$ & $21$ & $0.059$ & $79$ & $0.262$ & $21$ & $0.014$ & $79$ & $4.569$ \\
			\hline
			256 & $-$ & $-$ & $-$ & $-$ & $25$ & $0.134$ & $75$ & $0.628$ & $25$ & $0.093$ & $75$ & $9.120$ \\ 
			\hline
			384 & $-$ & $-$ & $-$ & $-$ & $20$ & $0.083$ & $80$ & $1.646$ & $20$ & $0.104$ & $80$ & $19.215$ \\
			\hline
			512 & $-$ & $-$ & $-$ & $-$ & $22$ & $0.143$ & $78$ & $2.131$ & $22$ & $0.112$ & $78$ & $29.048$ \\
			\hline
			768 & $-$ & $-$ & $-$ & $-$ & $19$ & $0.426$ & $81$ & $2.241$ & $19$ & $0.404$ & $81$ & $55.991$ \\
			\hline
			1024 & $-$ & $-$ & $-$ & $-$ & $17$ & $0.847$ & $83$ & $10.545$ & $17$ & $1.679$ & $83$ & $60.000$ \\
			\hline
			1536 & $-$ & $-$ & $-$ & $-$ & $16$ & $0.369$ & $84$ & $7.844$ & $16$ & $0.943$ & $84$ & $60.000$ \\
			\hline
			2048 & $-$ & $-$ & $-$ & $-$ & $15$ & $1.651$ & $85$ & $16.179$ & $14$ & $1.987$ & $86$ & $60.000$ \\
			\hline
			3072 & $-$ & $-$ & $-$ & $-$ & $21$ & $1.059$ & $79$ & $46.589$ & $21$ & $2.999$ & $79$ & $60.000$ \\
			\hline
			4096 & $-$ & $-$ & $-$ & $-$ & $17$ & $1.551$ & $78$ & $91.963$ & $18$ & $4.846$ & $82$ & $60.000$ \\
			\hline
		\end{tabular}
	}
\end{table}

The computational results for directed multigraphs (Table~\ref{table:random_directed}, Fig.~\ref{image:computational_results_graph_directed_solution} and~\ref{image:computational_results_graph_directed_no_solution}) turned out to be fundamentally different. Only about 20\% of random instances were feasible. Note that this does not mean that the corresponding vertices of the traveling salesperson polytope are adjacent since the second Hamiltonian decomposition problem only verifies a sufficient condition for the non-adjacency in a 1-skeleton.

The worst results of the three considered algorithms were shown by backtracking based on simple path extension (BSP).
In the allotted time, the algorithm solved only 225 instances out of 1500, without solving a single problem for graphs with more than 64 vertices. While chain edge fixing (BCEF) solved 1495 instances, and the general variable neighborhood search (GVNS) solved all instances. In terms of running time on feasible problems, the GVNS heuristic algorithm showed an advantage over BCEF on small graphs up to 512 vertices by an average of 6 times. However, on graphs with more than 512 vertices, BCEF turned out to be on average 2 times faster than GVNS. In general, for feasible problems, the BCEF and GVNS algorithms showed comparable results. Some variations in performance may be due to different implementations of the algorithms. 
However, we note that with an increase in the graph order, the advantage of BCEF over GVNS also increased.
On the other hand, the general variable neighborhood search heuristics (GVNS) has encountered significant difficulties in infeasible instances. The algorithm cannot determine this scenario and exits only when the limit on running time or number of iterations is reached. On such problems, the algorithm based on chain fixing of edges (BCEF) turned out to be on average 16 times faster than GVNS. In part, this means that the iteration threshold for the heuristic algorithm could be lowered. However, in this case, there would be a danger of losing existing solutions.

\section*{Conclusion}

In this paper, two backtracking algorithms were considered for the problem of constructing a second Hamiltonian decomposition of a 4-regular multigraph. According to the results of computational experiments on random directed and undirected multigraphs, the backtracking algorithm based on chain edge fixing turned out to be much more efficient than backtracking based on simple path extension. In addition, on directed multigraphs, the chain edge fixing showed comparable results with the previously known general variable neighborhood search heuristics on feasible instances and significantly outperformed the heuristics on infeasible problems.

The considered backtracking algorithms were developed for the problem of verifying the vertex non-adjacency in a 1-skeleton of a traveling salesperson polytope. However, they can also be applied directly to the problem of constructing the Hamiltonian decomposition of a regular multigraph and many of its applications.

\printbibliography

\end{document}